\documentclass[journal]{IEEEtran}

\usepackage{setspace}
\usepackage{cite,graphicx,amsmath,amssymb}
\usepackage{hyperref}
\usepackage{subfigure}
\usepackage{url}
\usepackage{fancyhdr}
\usepackage{mdwmath}
\usepackage{mdwtab}
\usepackage{balance}
\usepackage{xcolor}
\usepackage{bm}
\usepackage{amsthm}
\usepackage{threeparttable}
\usepackage{algorithm}
\usepackage{algorithmic}
\usepackage{multirow}
\usepackage{flafter}
\usepackage{makecell}
\usepackage{diagbox}
\usepackage{tcolorbox}
\theoremstyle{definition}

\newtheorem{theorem}{Theorem}

\newtheorem{lemma}{Lemma}

\newtheorem{corollary}{Corollary}

\usepackage{xurl}

\providecommand{\url}[1]{#1}
\allowdisplaybreaks
\begin{document}

\title{Simultaneously Transmitting And Reflecting (STAR) RIS for 6G: Fundamentals, Recent Advances, and Future Directions}

\author{Yuanwei Liu, Jiaqi Xu, Zhaolin Wang, Xidong Mu, Jianhua Zhang, and Ping Zhang\\

\thanks{Yuanwei Liu, Jiaqi Xu, Zhaolin Wang, and Xidong Mu are with the School of Electronic Engineering and Computer Science, Queen Mary University of London, London E1 4NS, UK, (email: \{yuanwei.liu, jiaqi.xu, zhaolin.wang, xidong.mu\}@qmul.ac.uk).}
\thanks{Jiahua Zhang and Ping Zhang are with the State Key Laboratory of Networking and Switching Technology, Beijing University of Posts and Telecommunications, Beijing 100876, China (email: \{jhzhang, pzhang\}@bupt.edu.com).}

}
\maketitle

\begin{abstract}
    Simultaneously transmitting and reflecting reconfigurable intelligent surfaces (STAR-RISs) have been attracting significant attention in both academia and industry for their advantages of achieving 360$^{\circ}$ coverage and enhanced degrees of freedom. This article first identifies the fundamentals of STAR-RIS, by discussing the hardware models, channel models, and signal models. Then, three representative categorizing approaches for STAR-RIS are introduced from phase-shift, directional, and energy consumption perspectives. Furthermore, the beamforming design of STAR-RIS is investigated for both independent and coupled phase-shift cases. A general optimization framework is proposed as the recent advances, which has high compatibility and provable optimality regardless of the application scenarios. As a further advance, several promising applications are discussed to demonstrate the potential benefits of applying STAR-RIS in the sixth-generation wireless network. Lastly, a few future directions and research opportunities are highlighted for motivating future work.
\end{abstract}

\section{Introduction} \label{sec:introduction}
Consisting of a large number of low-cost and low-power electromagnetic elements, reconfigurable intelligent surfaces (RISs) have emerged as a promising technology for the sixth-generation (6G) wireless networks and thus received significant research interests from both academia and industry \cite{ahead,9424177,9136592,8910627}. By deploying RISs in existing wireless environments, the phase shift and amplitude of the wireless signal incident upon each RIS element can be adjusted. Thus, the propagation of the reflected wireless signal can be changed for satisfying specific requirements \cite{ahead,9424177}. In contrast to wireless networks from the previous first-generation (1G) to the current fifth-generation (5G), where the wireless technologies were generally developed to combat the random radio environment (e.g., fadings and blockages), RISs achieve the so-called ``smart radio environment'' for realizing 6G in a flexible and sustainable \cite{ahead}. More importantly, their two-dimensional form factors and nearly passive working modes make RISs highly compatible with existing wireless technologies. The superiority of RISs has been extensively studied for achieving different objectives (e.g., spectral-/energy-efficiency enhancement \cite{8741198, 8982186, 9110912} and transmit power reduction \cite{8811733, 8970580, Xianghao}) and in diversified communication scenarios (e.g., multi-cell communications \cite{9090356, 9246254}, simultaneous information and power transfer (SWIPT) \cite{8941080, 9110849, 9148892}, physical layer security (PLS) \cite{8743496, 8723525, 9133130}, and unmanned aerial vehicles (UAVs) \cite{8959174, 9013626, 9277627}). 

\subsection{The Road to Simultaneously Transmitting And Reflecting (STAR)-RIS}
Note that most of the existing research contributions on RISs focused on the category of single-functional RISs, i.e., either transmitting or reflecting  incident wireless signals. This inevitably leads to a 180$^{\circ}$ smart radio environment. Therefore, wireless devices located on one side of a reflecting/transmitting-only RIS can merely benefit from the facilitated smart radio environment, while the other side of the RIS remains ineffective or possibly even worse than the conventional radio environment. As a remedy, a novel simultaneously transmitting and reflecting (STAR)-RIS technology was first proposed by \cite{xu_star}. As illustrated in Fig. \ref{f1}, STAR-RIS integrates both transmission and reflection (T\&R) functions. Notably, the wireless signal incident upon the STAR-RIS can be transmitted and reflected into both sides of the STAR-RIS with reconfigured propagation. For example, as illustrated, the incident signal can be steered into a given direction in both T\&R space. As a result, a 360$^{\circ}$ smart radio environment can be realized by STAR-RISs. Compared to conventional reflecting/transmitting-only RISs, there are two key benefits of STAR-RISs:
\begin{figure}[h!]
    \begin{center}
        \includegraphics[scale=0.8]{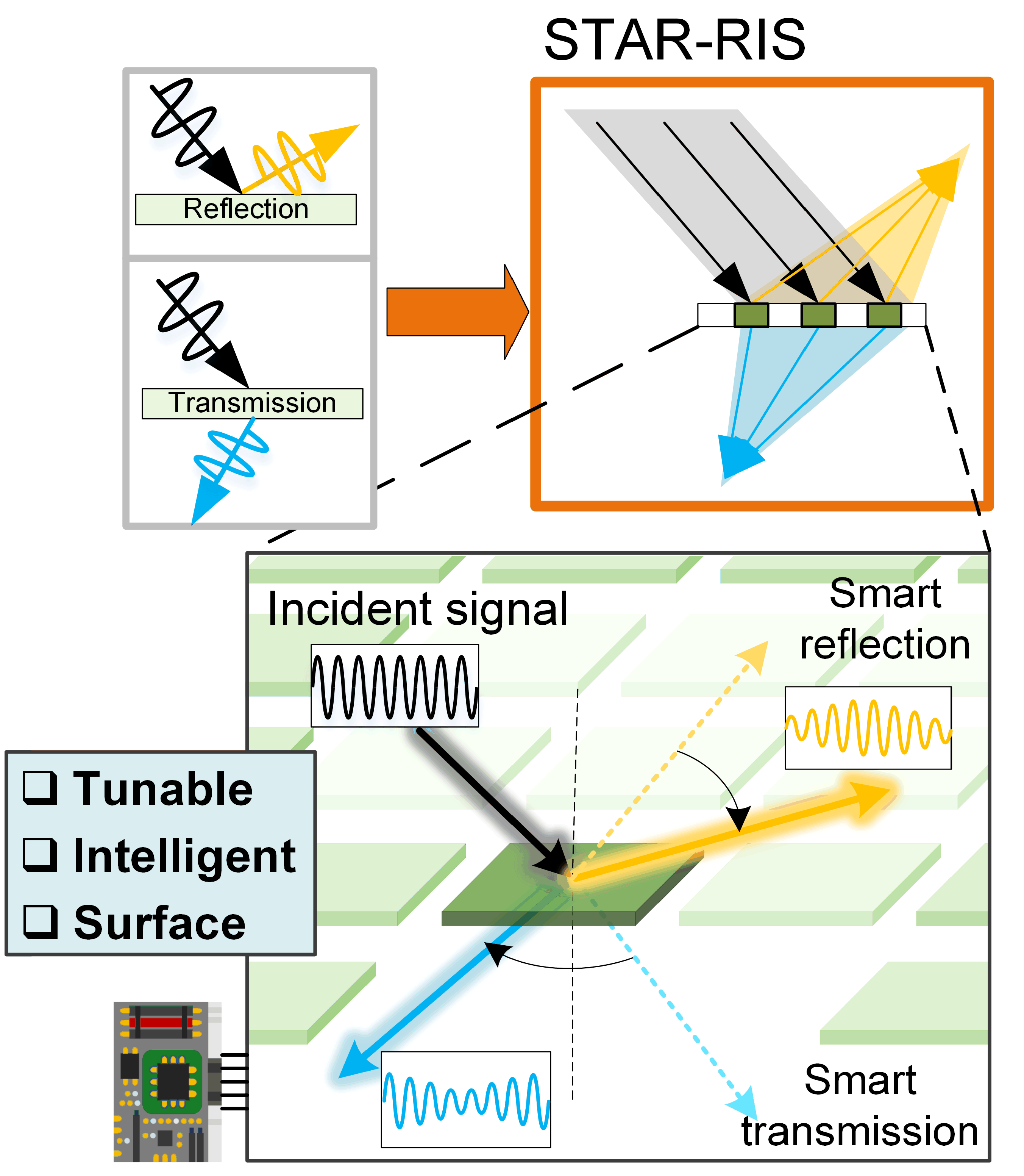}
        \caption{Illustration of the STAR-RIS and the facilitated 360$^{\circ}$ smart radio environment.}
        \label{f1}
    \end{center}
\end{figure}
\begin{itemize}
  \item \textbf{Full Space Coverage}: The realized 360$^{\circ}$ smart radio environment allows wireless nodes to receive the beneficially reconfigured wireless signals regardless of their locations with respect to the STAR-RIS. Therefore, no additional topological constraints have to be satisfied like conventional reflecting/transmitting-only RISs. STAR-RISs contribute to ubiquitous coverage for 6G with high flexibility.
  \item \textbf{Enhanced Degrees-of-Freedom (DoFs)}: As illustrated in Fig.~\ref{f1}, desired propagation can be achieved on both sides via the T\&R coefficients provided by STAR-RISs. Compared to conventional reflecting/transmitting-only RISs, more DoFs can be exploited in STAR-RIS assisted communications.
\end{itemize}

\subsection{Overview of Research Contributions on STAR-RIS}
Motivated by the uniqueness of STAR-RISs, extensive research efforts have been devoted. In the following, we focus on overviewing research contributions in terms of hardware and channel modeling, performance analysis, and beamforming optimization of STAR-RISs.

\subsubsection{Hardware and Channel Modeling}
Since there are various types of STAR-RIS, existing works proposed different hardware and signal models. \cite{9200683,zhang2020intelligent} proposed an initial signal model for STAR-RISs where the T\&R phase shifts are considered to be the same. \cite {xu_star} proposed a hardware model and near/far-field channel models for independent T\&R phase-shift STAR-RISs. \cite{a2} proposed a hardware implementation for metasurface-based STAR-RIS to achieve independent control of T\&R phase shifts. \cite{xu_vt} discussed different hardware implementations, hardware models, and channel models for STAR-RISs.

\subsubsection{Performance Analysis} 
Some other works also investigated STAR-RIS from the performance analysis perspective,
\cite{9837436} studied the performance of STAR-RIS-aided non-orthogonal multiple access (NOMA) networks with phase quantization errors and channel estimation errors. \cite{a16} investigated the performance of STAR-RIS-aided downlink NOMA networks with randomly deployed users. \cite{b27} investigated the performance of STAR-RIS-enabled multiple two-way full-duplex device-to-device communication systems over Rayleigh fading channels. \cite{9794569} studied the performance of a STAR-RIS-aided communication system considering the impacts of actual phase quantization errors and outdated channel state information. \cite{b52} studied the outage probability of a STAR-RIS-assisted downlink NOMA network over spatially correlated channels. \cite{c24} investigated the coverage probability of a STAR-RIS-assisted massive multiple-input multiple-output (mMIMO) system.

\subsubsection{Beamforming Optimization} 
To fully leverage the benefits of the STAR-RIS, it is essential to conduct a proper optimization of the T\&R coefficients, which have attracted substantial research interests. \cite{9685199} considered a two-user system, where the active beamforming, phase shifts of T\&R coefficients, and energy splitting ratios of T\&R coefficients are optimized to minimize the transmit power. Following this line of work, \cite{9920228} studied the beamforming design of STAR-RIS-aided communication systems in a more general multi-user scenario. Moreover, \cite{Mu_twc} made novel contributions by proposing three practical operating protocols for the STAR-RIS, namely energy splitting, mode switching, and time switching. Subsequently, joint optimization schemes were developed for these protocols to minimize transmit power. From a different perspective, \cite{9629335} maximized the weighted sum rate under the three operating protocols of STAR-RIS. The aforementioned studies assumed that the phase shifts of T\&R coefficients of STAR-RIS can be adjusted independently. However, in practical scenarios, those phase shifts may be coupled, thereby motivating recent investigations on beamforming design for the coupled phase-shift STAR-RIS. Specifically, \cite{Liu_ICC} developed a low-complexity element-wise algorithm for the coupled phase-shift STAR-RIS that served two communication users. \cite{niu2022weighted} considered a more general multi-user system, where the coupled phase shifts were optimized through alternating optimization. As a novel contribution, \cite{Zhaolin_wcl} proposed a generalized optimization framework for the coupled phase-shift STAR-RIS that has provable optimality under mild conditions. Additionally, \cite{9837935} employed hybrid reinforcement learning techniques to obtain the T\&R coefficients that comply with the coupled phase-shift constraints.

\subsection{Motivations and Organizations}
As can be observed, STAR-RIS is a prominent member of the RIS family and the research contributions on STAR-RISs keep rapidly growing. However, the development of STAR-RISs has been extensively researched in various application scenarios, perspectives, and assumptions without much coordination. There is an urgent need to categorize the existing studies, which provides the main motivation of this article. In line of this, this article intended to consolidate and advance the research on STAR-RISs via summarizing the fundamentals of STAR-RISs, overviewing the recent state-of-the-art research results on STAR-RISs, as well as highlighting promising applications and open problems for STAR-RISs.

The remainder of this paper is structured as follows. Section II introduces the fundamentals of STAR-RISs covering the hardware models, channel models, and signal models. Section III discusses three representative categorizing approaches for STAR-RISs from phase-shift, directional, and energy consumption perspectives. Section IV discusses the STAR beamforming design for the independent and coupled phase-shift models, followed by proposing a generalized optimization framework for STAR-RISs. Section V highlights several promising STAR-RIS enabled wireless technologies for 6G and provides numerical results for demonstrating the effectiveness. Conclusions are draw and some future directions for investigating STAR-RISs are identified in Section VI. 

\section{Fundamentals of STAR-RISs} \label{sec2}

In this section, we discuss the fundamentals of STAR-RIS in terms of hardware models, channel models, and signal models.
Firstly, hardware modeling studies how STAR-RIS interacts with wireless signals. Thus, relations between physical parameters and T\&R coefficients for each element can be obtained. Secondly, through channel modeling, we study channels between STAR-RISs and receivers so that end-to-end channel gains can be obtained. 
Finally, signal models can be established for determining the reconfigured signals by STAR-RISs.

\subsection{Hardware Models}\label{sec_hard}
The hardware implementations of STAR-RISs can be broadly classified into two categories, namely, \textit{patch-array-based STAR-RISs} and \textit{metasurface-based STAR-RISs} \cite{xu_vt}. Since STAR-RISs have a spatially periodic structure, the key difference between these two categories is that patch-array-based and metasurface-based STAR-RISs have significantly different period sizes.
\begin{itemize}
    \item \textbf{Patch-array-based STAR-RISs} \cite{9200683}: They consist of periodic elements having sizes on the order of a few centimetres. As result, the phase-shift profile of these STAR-RISs is spatially discrete.
    \item \textbf{Metasurface-based STAR-RISs} \cite{doc}: They have periodic cells on the order of a few millimetres down to molecular sizes. As a result, metasurface-based STAR-RISs are able to achieve the spatially-continuous configuration of incident waves.
\end{itemize}

In terms of hardware modeling, these two categories of STAR-RIS hardware implementations require different modeling methods. Specifically, the load impedance model is suitable for patch-array-based STAR-RISs and the generalized sheet transition conditions (GSTC)
model is suitable for modeling metasurface-based STAR-RISs. For the load impedance model, each element of a STAR-RIS can be treated as a lumped circuit with electric and magnetic impedances of $Y_m$ and $Z_m$. By configuring the values of two complex-valued impedances, the T\&R coefficients of the $m$th element can be controlled through the relation: $
T_m = \frac{2-\eta_0Y_m}{2+\eta_0Y_m}-R_m$, and $R_m = -\frac{2(\eta^2_0Y_m-Z_m)}{(2+\eta_0Y_m)(2\eta_0+Z_m)}$,\vspace{0.06in} where $\eta_0$ is the impedance of free space \cite{xu_star}. For the GSTC model, metasurface-based STAR-RISs are characterized by electric and magnetic susceptibilities as functions of position \cite{ahead}. The discontinuity of electric and magnetic fields on the two sides of the metasurface can be obtained through these susceptibilities. Compared with the load impedance model, the GSTC model is able to obtain a more detailed result for the distribution of the transmitted and reflected signals.

\subsection{Channel Models}
For STAR-RIS, different channel modeling methods are required for communication in far-field and near-field regions. A wireless receiver is regarded to locate in the far-field region of a STAR-RIS if the communication distance $d > 2L^2/\lambda$, where $L$ is the largest dimension of STAR-RIS and $\lambda$ is the carrier wavelength \cite{ahead}. 
\begin{itemize}
    \item \textbf{Far-field Channel Models}: Within the far-field, the Friis formula is valid for free-space transmission and the received power falls of with $d_1^{-2}d_2^{-2}$, where $d_1$ is the communication distance between the transmitter and STAR-RIS and $d_2$ is the distance between STAR-RIS and a receiver.
The overall channel between a transmitter and a receiver through STAR-RIS is a cascaded link where the link through each element is scaled with the corresponding T\&R coefficients.
In the presence of scatters, the STAR-RIS-aided far-field channel can be modeled by exploiting the Rician distribution.

\item \textbf{Near-field Channel Models}: Near-field channel models involve not only the distance $d_2$, but also all the distances between each STAR-RIS element and a receiver. The near-field channel can be formulated as the sum of all these individual links with different path-lengths. In addition, for metasurface-based STAR-RIS, the near-field channels are modeled through the Huygens-Fresnel principle-based integration \cite{xu_star}. For these cases, the near-field channel is an integral over the surface area of the STAR-RIS, where contributions of each unit area are proportional to its size, its amplitude, its leaning factor, and its reciprocal distance to the receiver.
\end{itemize}

\subsection{Signal Models} \label{sub_signal} 
Here, we provide basic signal models for STAR-RISs. In most cases, wireless signals incident on a given element of the STAR-RIS are divided into transmitted and reflected signals. Let $s_m$ denote the signal incident on the one side of the $m$th STAR element. Thus, transmitted signals and the reflected signals can be modeled as $t_m = (\sqrt{\beta^t_m}e^{j\phi^t_m})\cdot s_m$ and $r_m = (\sqrt{\beta^r_m}e^{j\phi^r_m})\cdot s_m$, respectively, where $\sqrt{\beta^{t/r}_m}\in [0,1]$ and $\phi^{t/r}_m \in [0,2\pi )$ denote the amplitude and phase shift of the T\&R coefficients. For different categories of STAR-RISs, there are in general additional constraints on amplitude and phase shift of the T\&R coefficients, which will be further discussed in the next section. For example, if the STAR element is considered to be lossless, energy conservation dictates: $|t_m|^2 + |r_m|^2 = |s_m|^2$. Equivalently, the amplitude of T\&R coefficients should satisfy: $\beta^r_m + \beta^t_m = 1$.

\section{Representative Categorizing Approaches of STAR-RISs}

In this section, we distinguish representative categorizing approaches of STAR-RISs from three perspectives. We also reveal the key differences between these STAR-RIS categories in terms of signal models.

\subsection{A Phase-Shift Perspective: Independent Versus Coupled Phase-Shift}
As discussed in \ref{sub_signal}, in the basic STAR-RIS signal model, the involved transmission phase shift and reflection phase shift can take values ranging from $0$ to $2\pi$. In the existing literature, there are two main categories for T\&R phase-shift reconfiguration, namely \textit{independent phase-shift model} and \textit{coupled phase-shift model}.
\begin{itemize}
    \item \textbf{Independent Phase-Shift Model}: The transmission phase shift ($\phi^t_m$) and the reflection phase shift ($\phi^r_m$) can be adjusted independently with each other. This independent phase-shift control is achieved by tuning surface electric and magnetic impedances. Recall that in \ref{sec_hard}, T\&R phase-shifts are determined through two complex-valued impedances, $Y_m$ and $Z_m$. In other words, each STAR element has four real-valued configurable DoFs which correspond to $\beta^t_m$, $\beta^r_m$, $\phi^t_m$, and $\phi^r_m$. However, the independent phase-shift model is challenging to realize in practice, especially for passive STAR-RISs. This is because if STAR-RISs are made of passive materials, the corresponding electric impedance and magnetic impedance cannot be arbitrary values. 
    \item \textbf{Coupled Phase-Shift Model}: As pointed out by \cite{zhu2014dynamic}, for STAR-RISs using passive lossless materials, the corresponding electric impedance and magnetic impedances should be purely imaginary numbers. Under this constraint, transmission phase shift ($\phi^t_m$) and reflection phase shift ($\phi^r_m$) are coupled subject to specific values of phase-shift differences as follows \cite{zhu2014dynamic,xu_correlated}.
    \begin{align}\label{pha}
    \phi^r_m - \phi^t_m = \frac{\pi}{2} + \nu_m\pi, \ \nu_m=0\ \text{or}\ 1, \ \nonumber \\ \forall m=1,2,\cdots,M,
    \end{align}
    where $\nu_m$ is referred to as the \textit{auxiliary bit} and can only take on two values, 0 or 1. In other words, i.e., the phase difference $\phi^r_m$ and $\phi^t_m$ is either $\pi/2$ or $3\pi/2$. 
\end{itemize}

\subsection{A Directional Perspective: Single-Sided STAR-RISs versus Dual-Sided STAR-RISs}
\begin{figure}[t!]
    \begin{center}
        \includegraphics[scale=0.35]{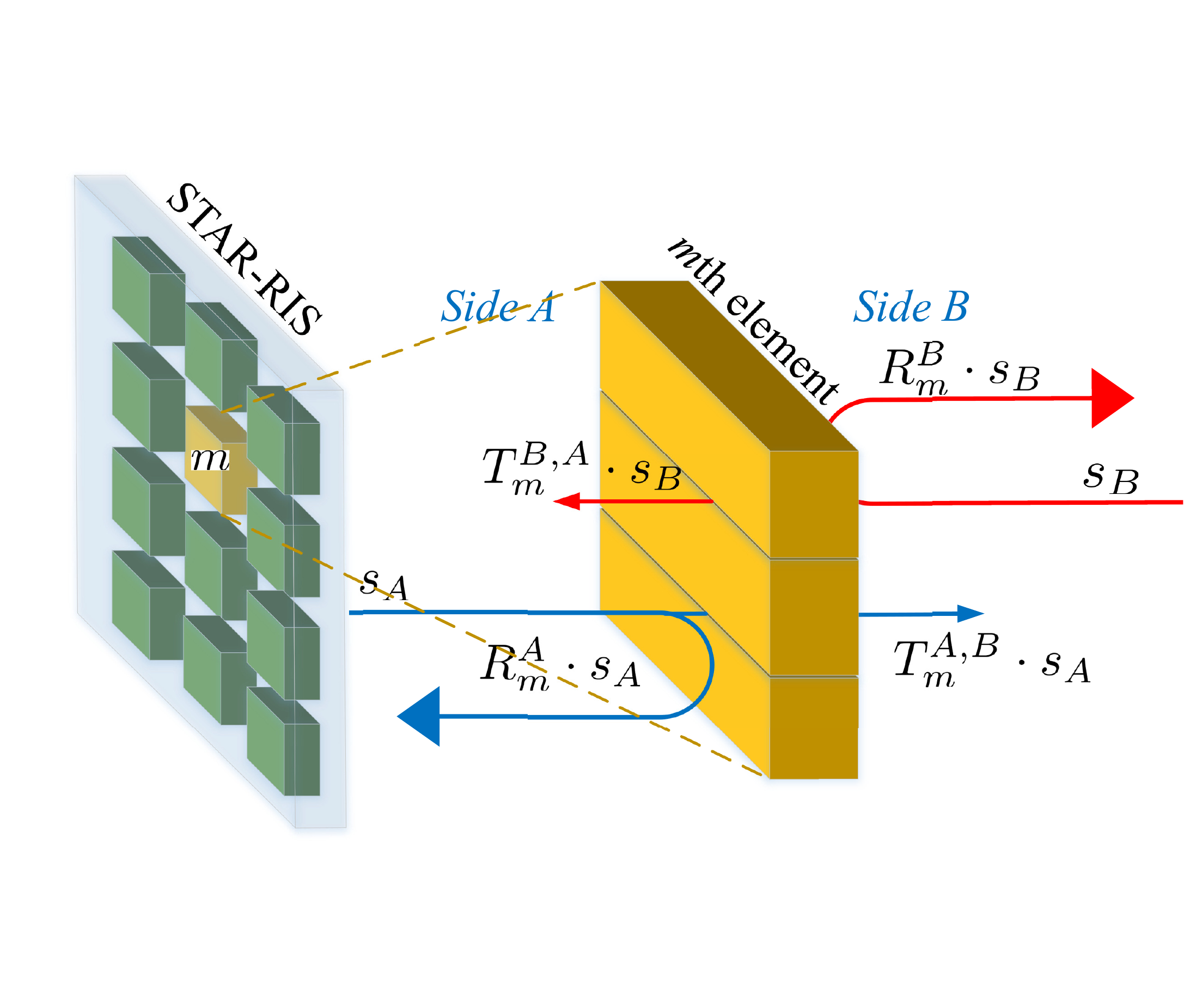}
        \caption{A general signal model for dual-sided STAR-RISs}
        \label{randt}
    \end{center}
\end{figure}
The basic signal model for STAR-RIS given in \ref{sub_signal} only considers signals incident from one side of the surface. 
In practice, wireless signals will incident upon both sides of STAR-RISs simultaneously.  
However, it is not yet clear how these signals can be simultaneously manipulated. To address this issue, \cite{9935303} proposed a signal model for \textbf{dual-sided STAR-RIS}. As illustrated in Fig.~\ref{randt}, under dual-sided incidence, each STAR element has four T\&R coefficients, i.e., $\Tilde{R}^A_m$, $\Tilde{R}^B_m$, $\Tilde{T}^{AB}_m$, and $\Tilde{R}^{BA}_m$. These four coefficients transform incident signals to output signals of each STAR element. The relation between incident and output signals is given by:

\begin{equation}\label{tr_def}
    \begin{pmatrix}
y^A_m\\
y^B_m
\end{pmatrix} =
    \begin{pmatrix}
\Tilde{R}^A_m & \Tilde{T}^{AB}_m\\
\Tilde{T}^{BA}_m & \Tilde{R}^B_m
\end{pmatrix}
\begin{pmatrix}
s^A_m\\
s^B_m
\end{pmatrix},
\end{equation}
or equivalently, $\mathbf{y}_m = \mathbf{\Xi}_m \cdot \mathbf{s}_m$, where $\mathbf{\Xi}_m$ is the T\&R matrix of the $m$th STAR element. In terms of reciprocity, a STAR-RIS can be either reciprocal or nonreciprocal. A STAR element is reciprocal if the T\&R matrix $\mathbf{\Xi}_m$ is symmetrical, i.e., $\Tilde{T}^{AB}_m = \Tilde{T}^{BA}_m$. For nonreciprocal STAR matrix, we have $\Tilde{T}^{AB}_m \neq \Tilde{T}^{BA}_m$. 
For application scenarios where all users under service are located on one side of the STAR-RIS (for example, \textit{side A}), signal power is lost if $\Tilde{T}^{BA}_m \neq 0$ even if all STAR elements might be lossless. An nonreciprocal STAR element is ideal in these scenarios to avoid power leakage from \textit{side A} to \textit{side B}.

\subsection{An Energy Consumption Perspective: Passive STAR-RISs versus Active STAR-RISs}

In terms of energy consumption, a patch-array-based STAR-RIS can be either passive or active.

\begin{itemize}
    \item \textbf{Passive STAR-RISs}: Since STAR-RISs generally do not require active components, most STAR-RIS implementations can be regard as passive. A STAR element is passive lossless if $\mathbf{\Xi}_m$ is unitary, i.e., $\mathbf{\Xi}_m^H\mathbf{\Xi}_m = \mathbf{I}_2$, where $\mathbf{I}_2$ denotes the two-by-two identity matrix. For passive lossless STAR elements, there is no energy loss so that the sum of reflected and transmitted energy is equal to the energy of incident signal. For passive lossy STAR element, a portion of incident energy is dissipated as heat, so that $\mathbf{\Xi}_m^H\mathbf{\Xi}_m \prec \mathbf{I}_2$. 
    \item \textbf{Active STAR-RISs}: For active STAR element, we have $\mathbf{\Xi}_m^H\mathbf{\Xi}_m \succ \mathbf{I}_2$ \cite{pozar2011microwave}. 
In terms of T\&R coefficient, the key difference between active and passive STAR-RISs is that for active STAR-RISs, the amplitudes of their T\&R coefficient can be greater than one, i.e., $\beta^r_m > 1$ and $\beta^t_m > 1$. This removes the amplitude constraint introduced for passive lossless STAR-RIS in \ref{sub_signal}.
As proposed by \cite{xu2023active}, active STAR-RISs can be achieved by exploiting reflection-type amplifiers and quadrature hybrid couplers. In addition, the performance gain of active STAR-RIS over passive ones is more significant when the number of STAR elements is small and the transmit signal-to-noise ratio (SNR) is low. 
\end{itemize}

\section{Beamforming Design for STAR-RISs: From Independent to Coupled Phase-Shift Models}
When deploying STAR-RISs to assist the communication systems, it is essential to carefully design the T\&R coefficients of each element, namely STAR beamforming design, for facilitating the desired signal propagation. In this section, we first introduce three practical operating protocols proposed for employing STAR-RISs. Then, we discuss the STAR beamforming approaches developed for both independent and coupled phase-shift models, and summarize their respective advantages and disadvantages. Finally, we propose a generalized optimization framework for STAR beamforming design, which has high compatibility and provable optimality.
\subsection{Operating Protocols}
As discussed in the previous section, there is a fundamental energy conservation constraint between the incident signal, transmitted signal, and reflected signal of each STAR element. As a result, each STAR can work in the intermediate mode, i.e., simultaneous transmission and reflection mode, and two extreme states, i.e., transmission-only and reflection-only modes. By exploiting this feature, three practical operating protocols were proposed by \cite{STAR_mag}, namely energy splitting, mode switching, and time switching, as follows.
\begin{itemize}
  \item \textbf{Energy Splitting}: All STAR elements work in the simultaneous transmission and reflection mode. Therefore, the signal incident upon each STAR element is generally split into transmitted and reflected signals towards different spaces. It can be observed that energy splitting STAR-RISs can exploit the maximum DoFs since the T\&R coefficients of each element can be optimized. This, however, leads to potentially high complexity in communication design and channel estimation given the resulting large dimension of optimization variables. Moreover, if the coupled phase-shift model is employed, the energy splitting STAR beamforming design more challenging than the case with the independent phase-shift mode.
  \item \textbf{Mode Switching}: All STAR elements work in either the transmission-only mode or the reflection-only mode. Note that since each element of mode switching STAR-RISs is only able to work in the single mode, the constraint caused by the coupled phase-shift model becomes a dummy. Therefore, the binary working mode reduces the total number of optimization variables and alleviates the overhead burden for controlling STAR-RISs, which is friendly for practical implementation. However, the maximum performance gain achieved by mode switching STAR-RISs is degraded due to the reduced DoFs.
  \item \textbf{Time Switching}: All STAR elements periodically work in the transmission-only mode and the reflection-only mode over different time slots. This leads to two independent T\&R coefficient designs, where the impact of the coupled phase-shift model also vanishes. Note that compared to mode switching STAR-RISs, the maximum array gain for transmission and reflection reserves in time switching STAR-RISs. However, it is also worth noting that time switching STAR-RISs have to frequently change the working modes of all elements, which leads to severe overhead and high implementation complexity.
\end{itemize} 
\subsection{STAR Beamforming Design under Independent and Coupled Phase-Shift Models} 
After introducing the operating protocols of STAR-RISs, we focus our attention on the fundamental STAR beamforming design. It can be observed that different operating protocols impose different constraints on the STAR beamforming design problem. As discussed before, compared to mode switching and time switching STAR-RISs, the beamforming design in energy splitting STAR-RISs is much more challenging. The case becomes even worse when considering the coupled phase-shift model. In the following, we first consider the energy splitting STAR beamforming design under the independent phase-shift model, where a joint optimization (JO) based optimization approach is introduced. Then, we consider the more challenging energy splitting STAR beamforming design under the coupled phase-shift model, where an effective element-wise optimization approach is introduced.
\begin{figure}[t!]
  \centering
  \includegraphics[width=2.8in]{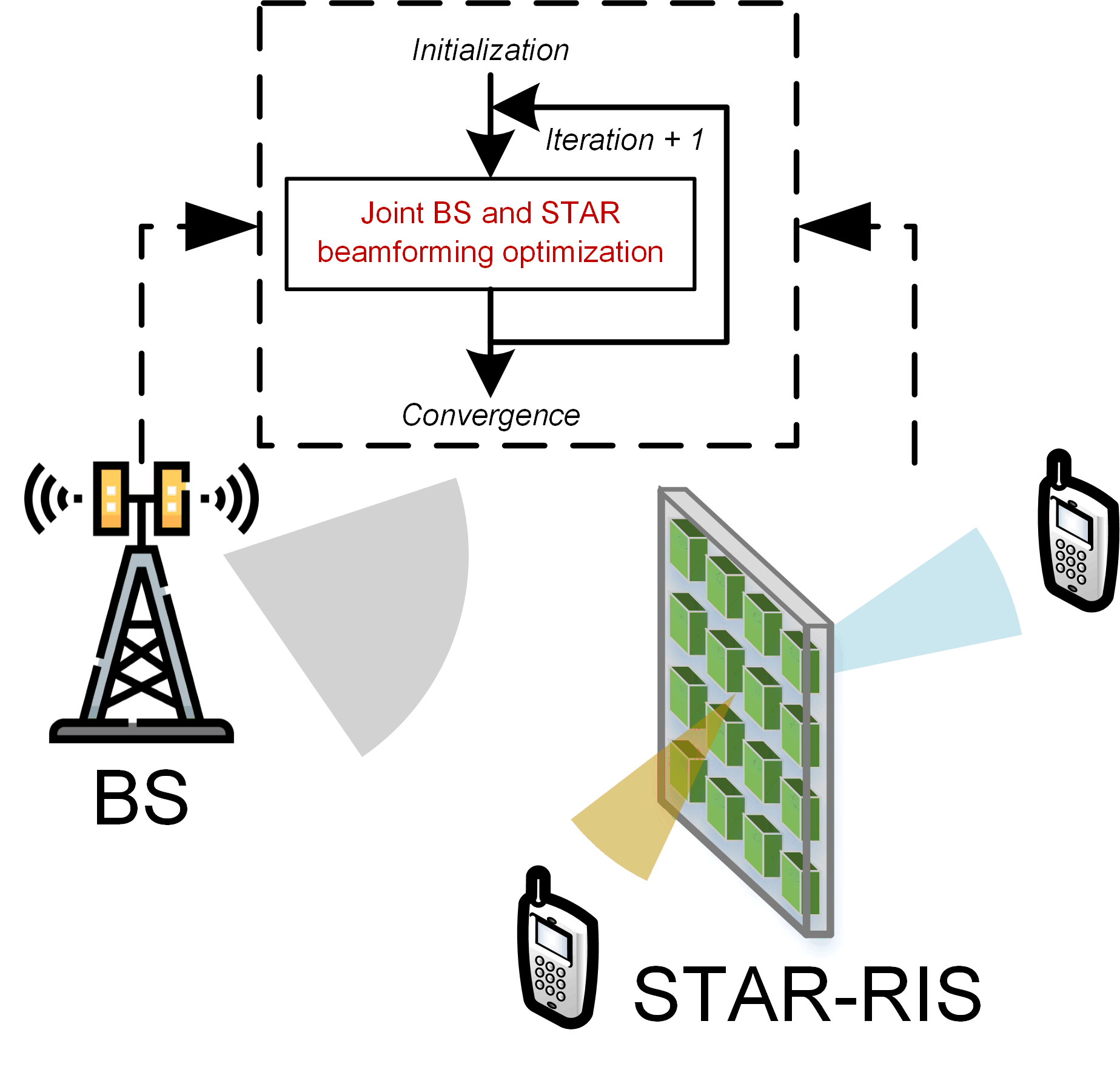}\\
  \caption{Joint Optimization Based STAR Beamforming Design.}\label{Independent}
\end{figure}

\begin{figure}[t!]
  \centering
  \includegraphics[width=3in]{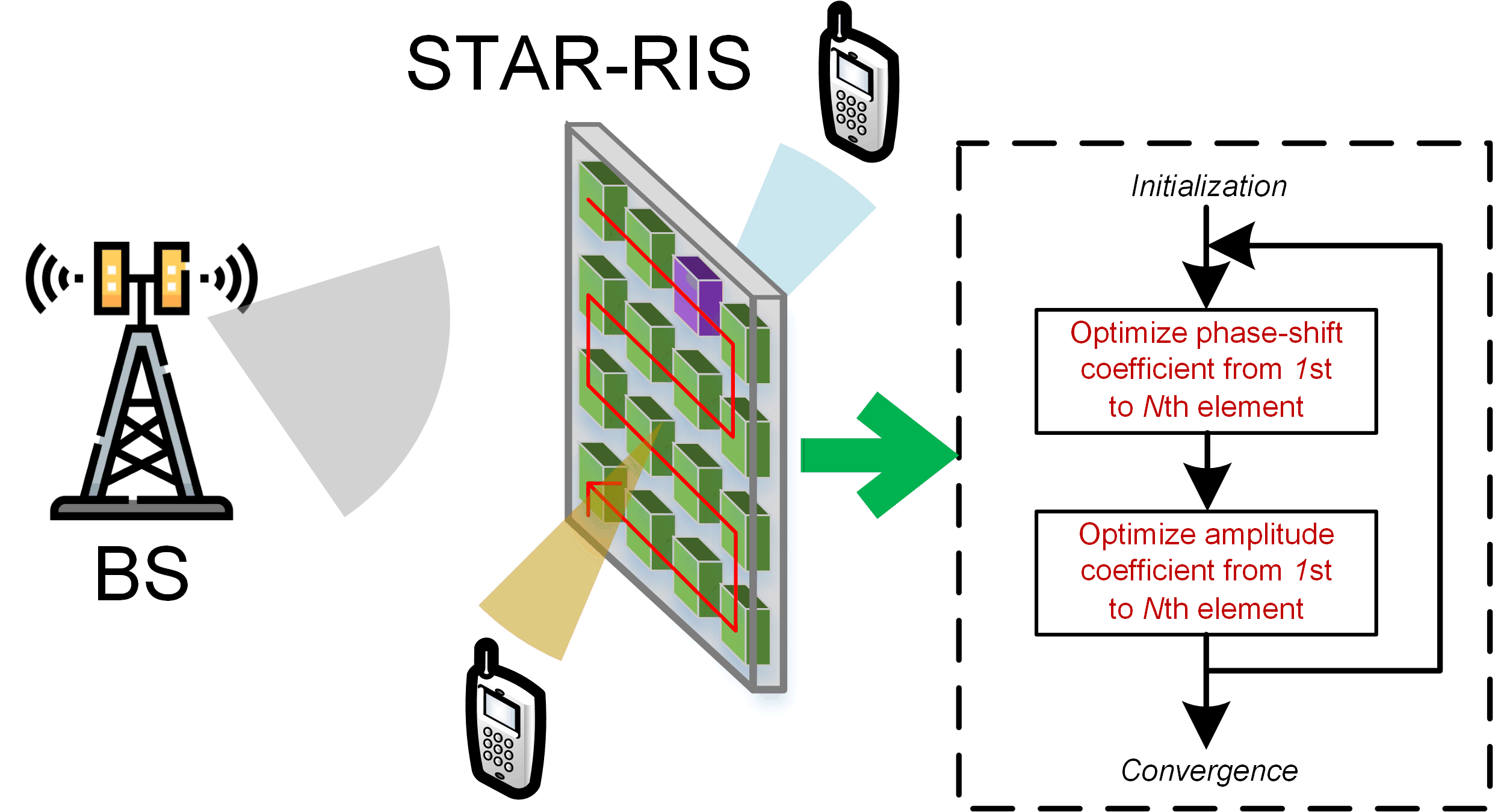}\\
  \caption{Element-Wise Based STAR Beamforming Design.}\label{Coupled}
\end{figure}

\subsubsection{Joint Optimization Based STAR Beamforming Design under the Independent Phase-Shift Model} For energy splitting STAR-RIS assisted communication systems with the independent phase-shift model, the coupling between the STAR beamforming only exists in the energy conservation constraint, which is or can be transformed into a convex constraint. Then, the main challenge lies in the coupling between the newly introduced STAR beamforming and the existing variables (e.g., base station (BS) beamforming and power allocation). As investigated in many existing RIS works, one straightforward way is to employ alternating optimization (AO), which decomposes the original joint BS and RIS beamforming problem into two subproblems and alternatingly optimizes one beamforming with the other fixed. The advantage of AO is that compared to the original problem, the two decomposed subproblems are less complicated and easy to be resolved. However, the main drawbacks of AO are that its convergence and optimality might not be guaranteed \cite{Xianghao}. To address this issue, \cite{Mu_twc} proposed an efficient JO based beamforming approach for minimizing the BS power consumption in a STAR-RIS assisted multiple-input single-output (MISO) multi-user communication system. As shown in Fig. \ref{Independent}, the BS beamforming and STAR beamforming in the proposed JO based approach can be simultaneously optimized in each iteration, thus guaranteeing convergence and obtaining a locally optimal solution. %The simulation code for the proposed JO based STAR beamforming is avaible at \url{https://github.com/STAR-Yuanwei-Liu/Optimization-for-wireless-communications}. 

\subsubsection{Element-Wise Based STAR Beamforming Design under the Coupled Phase-Shift Model} For energy splitting STAR-RIS assisted communication systems with the coupled phase-shift model, the STAR beamforming design becomes quite challenging. This is because for each STAR element, T\&R phase shifts are coupled, which leads to non-convex constraints. To overcome this obstacle, \cite{Liu_ICC} proposed an efficient element-wise based STAR beamforming approach for minimizing the BS power consumption in a STAR-RIS assisted single-input single-output (SISO) two-user communication system. As shown in Fig.~\ref{Coupled}, the salient feature is that the phase-shift and amplitude coefficients of each STAR element are optimized one by one, i.e., in an element-wise manner. The advantage is that the computational complexity only linearly scales with the number of STAR elements, which makes it promising to be used in practice since the size of STAR-RISs is usually large. Nevertheless, the limitations of the proposed element-wise based approach are that it can only be employed in the two-user communication system and there is no knowledge of the optimality of the obtained solution. %The simulation code for the proposed element-wise based STAR beamforming is avaible at \url{https://github.com/STAR-Yuanwei-Liu/Optimization-for-wireless-communications}.  

\subsection{A Generalized Optimization Framework for STAR Beamforming Design}
Having introduced two representative STAR beamforming approaches, we can observe that they still have the following two common drawbacks. 1) These approaches are in general developed for specific objective functions (e.g., power consumption and sum rate) and/or communication scenarios (e.g., MISO, SISO, and two-user cases). Their applicability to other use cases is limited. 2) The convergence of these approaches and the optimality of their solutions might not be always stable, especially for the coupled phase-shift case.

To fill in this gap, \cite{Zhaolin_wcl} proposed a generalized penalty-based optimization framework for STAR beamforming. The beauty of the proposed framework can be summarized as follows:
\begin{itemize}
  \item \textbf{High Compatibility}: The applications of the proposed framework are not restricted by objective functions and communication scenarios. More importantly, it can be regarded as an ``add-on'' approach to existing state-of-the-art approaches, thus being applicable to diversified STAR-RIS assisted communication scenarios.
  \item \textbf{Provable Optimality}: The solutions obtained by the proposed framework can be proved to be Karush-Kuhn-Tucker optimal under some mild conditions, such as the Robinson's condition or the Mangasarian-Fromovitz constraint qualification (MFCQ) condition. 
  \item \textbf{Guaranteed Convergence}: The convergence of the proposed framework can always be achieved regardless of the optimality of the obtained solutions.
\end{itemize} 
The simulation codes for the JO based approach, the element-wise based approach, and the generalized optimization framework for STAR beamforming design are available at \url{https://github.com/STAR-Yuanwei-Liu/Optimization-for-wireless-communications}.  

\begin{figure}[t!]
    \centering
    \subfigure{
        \label{coupled_convergence}
        \includegraphics[width=0.8\linewidth]{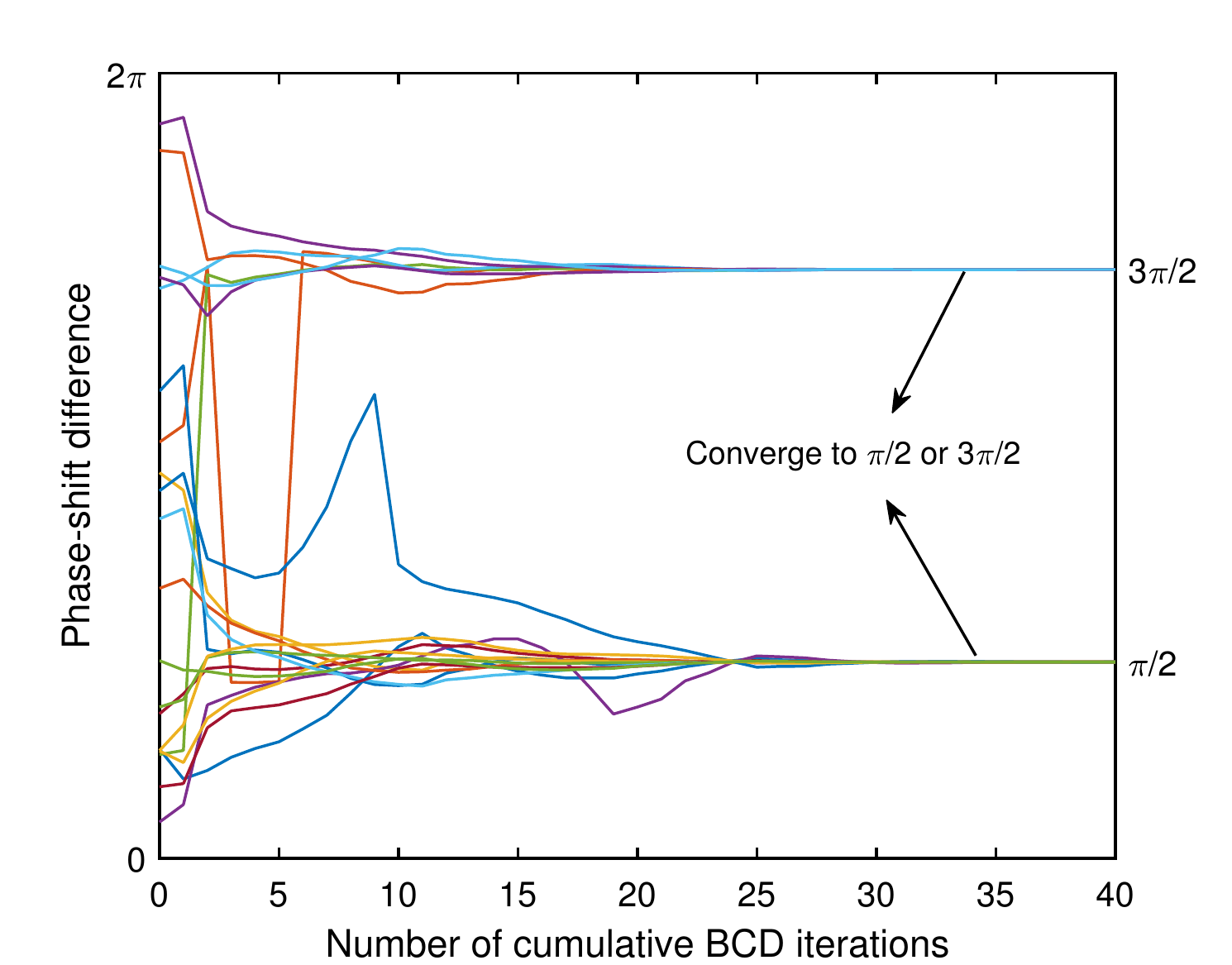}
    }
    \subfigure{
        \label{coupled_SE}
        \includegraphics[width=0.8\linewidth]{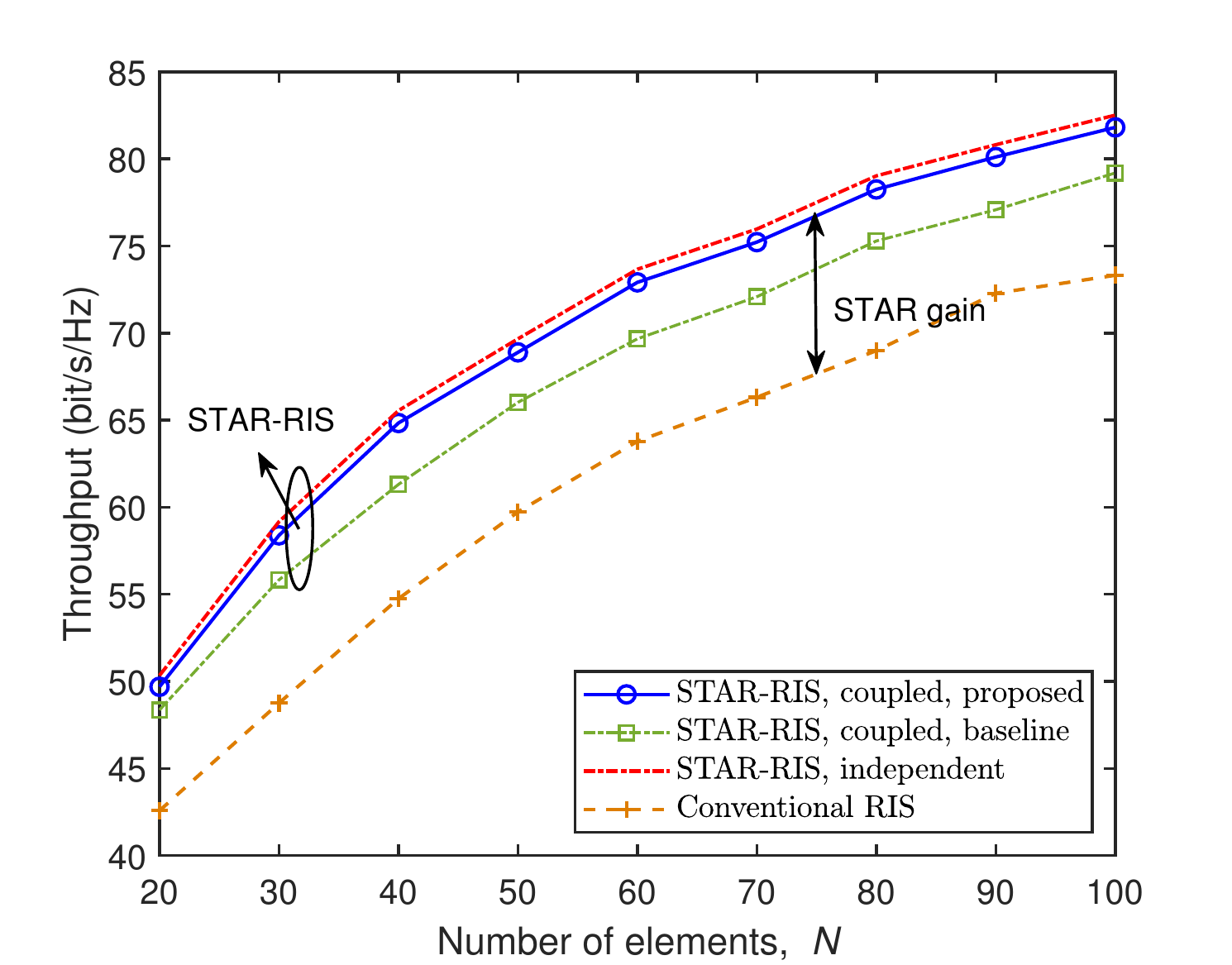}
    }
    \caption{Performance of the proposed generalized penalty-based optimization framework for maximizing the system spectral efficiency. a) The convergence performance of the phase-shift difference between the T\&R coefficients of each element of the STAR-RIS. b) The performance of the proposed framework in terms of spectral efficiency maximization, where “STAR-RIS, coupled, baseline” refers to the alternating optimization scheme \cite{niu2022weighted}. The detailed simulation setup can be found in \cite{Zhaolin_wcl}}
    \label{coupled}
\end{figure}

To further illustrate the efficacy of the generalized optimization framework in addressing STAR beamforming design problems, we present a case study involving a spectral efficiency maximization problem, as depicted in Fig.~\ref{coupled}. As illustrated in Fig.~\ref{coupled}a, the phase-shift differences between $\phi_m^r$ and $\phi_m^t$ for each element eventually converge to either $\frac{\pi}{2}$ or $\frac{3\pi}{2}$, thereby satisfying the coupled phase-shift constraint. Furthermore, as evidenced in Fig~\ref{coupled}b, the proposed framework achieves comparable performance to the independent phase-shift STAR-RIS approach and outperforms the existing baseline scheme based on alternating optimization \cite{niu2022weighted}, which confirms the optimality of the proposed framework.

\section{STAR-RIS enabled Promising Wireless Technologies towards 6G}
In this section, we continue to discuss recent advances in employing STAR-RIS to improve the performance of other 6G wireless technologies, including NOMA, integrated sensing and communication (ISAC), terahertz (THz) communications, and PLS. Numerical results are also provided to demonstrate the great potential of STAR-RISs in 6G.   

\subsection{STAR-RIS-NOMA}
RISs are capable of enhancing the performance of NOMA networks by providing distinct channel conditions for NOMA users \cite{9801736}. Nevertheless, for RIS-aided NOMA networks, all the users have to be set in the reflection space. As a result, the channel conditions of users in reflected space are generally similar, which is not easy to fully exploit the benefits of NOMA. STAR-RISs can properly solve this issue. More particularly, a pair of users at the transmission and reflection side can be grouped together for facilitating NOMA. As a result, asymmetric channel conditions among transmit and reflect users can be realized and thus high-performance gain over orthogonal multiple access (OMA) can be achieved by adopting flexible resource allocation. 

\cite{9863732} first proposed the framework for applying NOMA to STAR-RIS networks. The cluster-based beamforming design was adopted in the proposed framework. We jointly optimized the decoding order, power allocation coefficients and beamforming at both the BS side and the STAR-RIS side. Compared to RIS-NOMA, We illustrated the advantages of the proposed STAR-RIS-NOMA framework via simulation results. There are still several research opportunities and challenges in STAR-RIS-NOMA design based on our proposed framework, for example, 1) the appearance of STAR-RIS makes the joint user assignment and beamforming design challenging, how to efficiently align and allocate different NOMA users into suitable beams and to provide proper optimization for mitigating intra-cluster and inter-cluster interference requires further study. 2) Another unsolved problem is facilitating coupled phase-shift design in the proposed STAR-RIS-NOMA framework. It is worth mentioning that with the aid of our proposed optimization framework \cite{Zhaolin_wcl}, this issue can be solved in a proper way.

\begin{figure}[t!]
    \centering
    \subfigure{
        \label{ISAC_full}
        \includegraphics[width=0.8\linewidth]{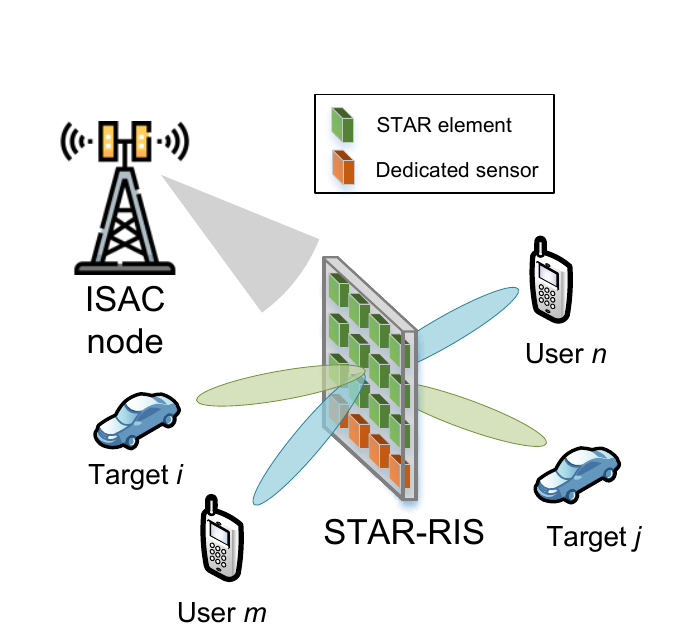}
    }
    \subfigure{
        \label{ISAC_half}
        \includegraphics[width=0.8\linewidth]{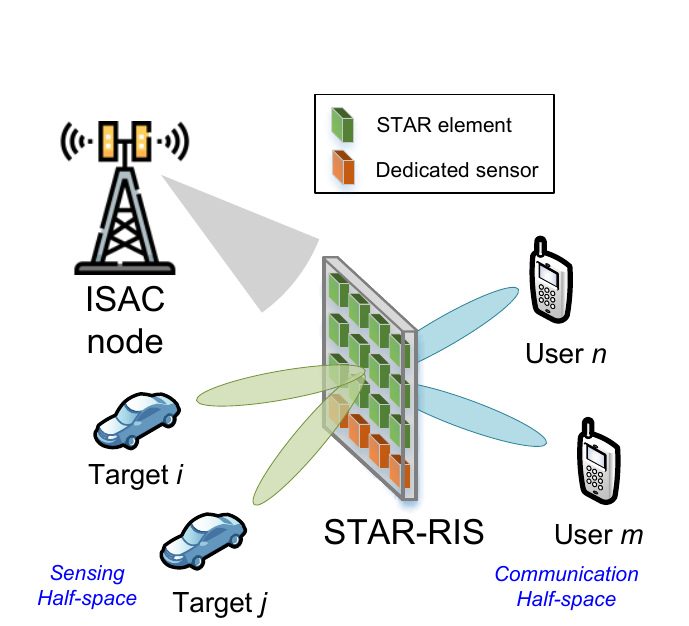}
    }
    \caption{Illustration of STAR-RIS-ISAC system for a) full-space sensing and communication and b) half-space sensing and communication.}
    \label{ISAC_model}
\end{figure}

\begin{figure}[t!]
    \centering
    \subfigure{
        \label{ISAC_tradeoff}
        \includegraphics[width=0.8\linewidth]{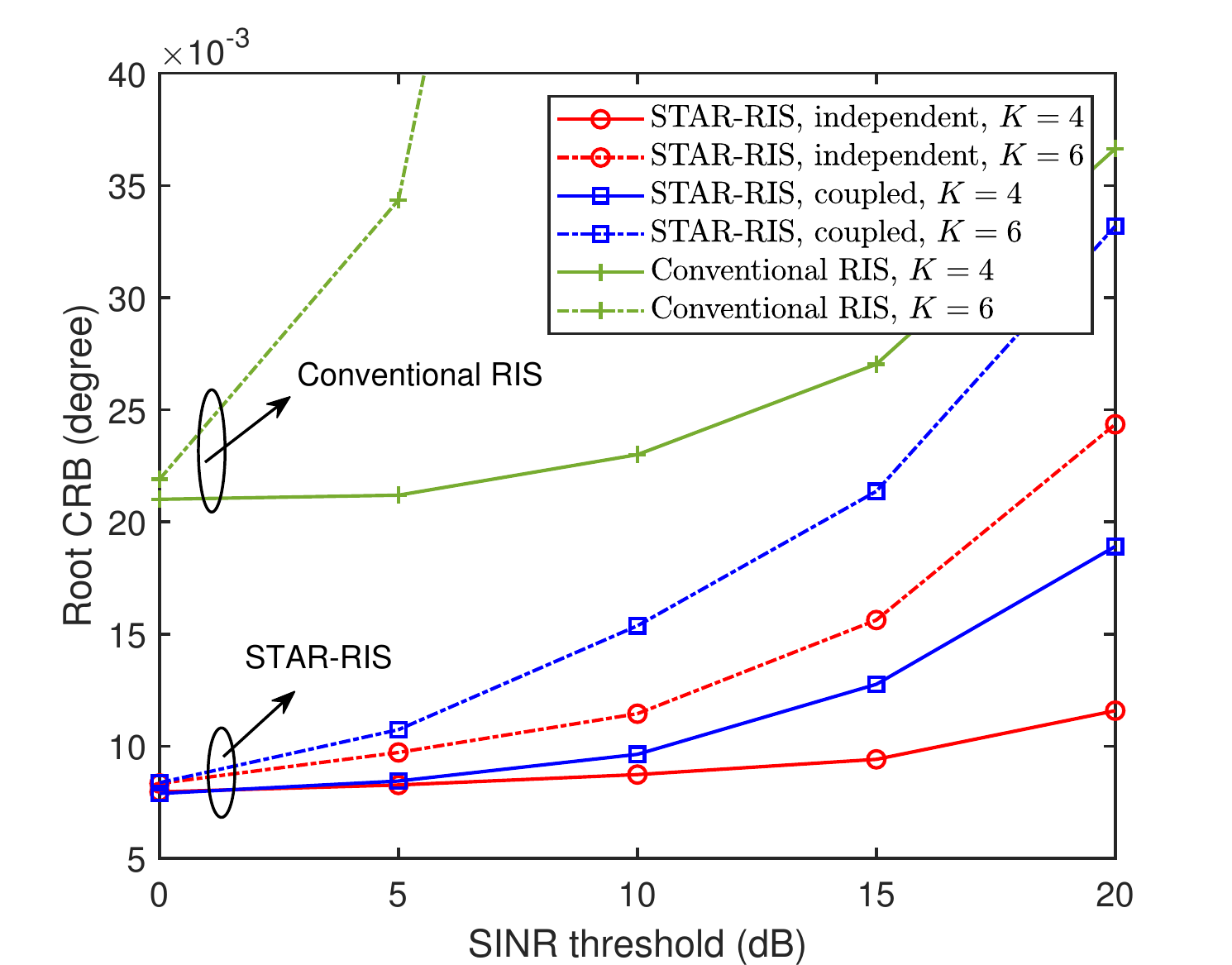}
    }
    \subfigure{
        \label{ISAC_MLE}
        \includegraphics[width=0.8\linewidth]{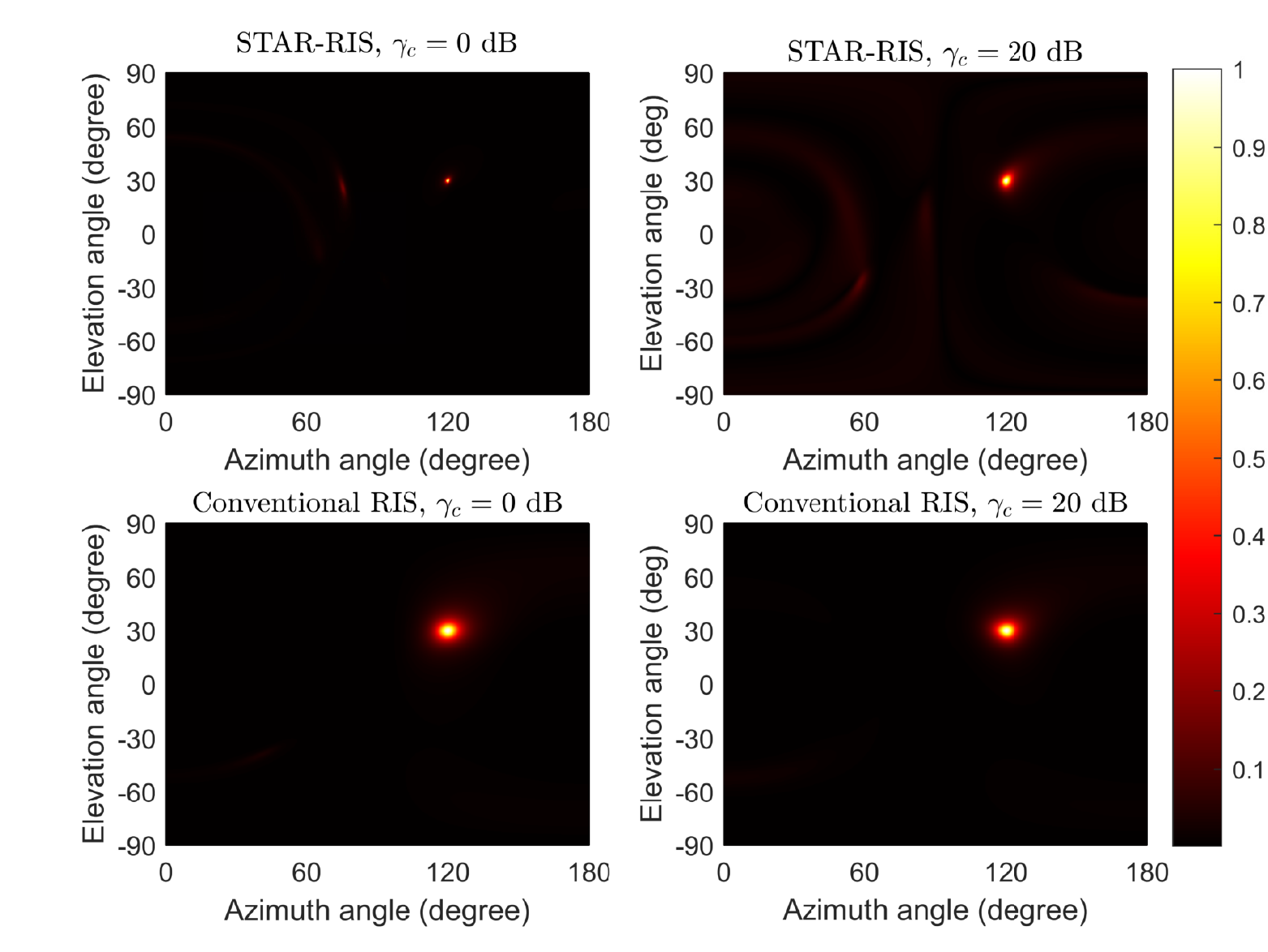}
    }
    \caption{Performance of the STAR-RIS-ISAC system with the \emph{sensing-at-STAR-RIS} structure. a) The achieved root Cramér–Rao bound (CRB) of target sensing under different communication SINR requirements, $\gamma_c$, where $K$ denotes the number of communication users. b) The sensing results obtained by the maximum likelihood estimation (MLE), where the brightest part represents the estimated location of the target. The detailed simulation setup can be found in \cite{wang2022stars}.}
    \label{ISAC}
\end{figure}

\subsection{STAR-RIS-ISAC}
ISAC is a technology that serves the dual purpose of sensing and communication through a shared hardware platform and waveform. Typically, ISAC relies on the line-of-sight (LOS) channel to facilitate the sensing function between the ISAC transceiver and the desired targets. Carrying out target sensing without LOS links presents a significant challenge. However, RIS can overcome this issue by establishing virtual LOS links between the RIS and  desired targets. This capability enhances the sensing coverage and resolution. Compared to conventional reflecting-only RIS, STAR-RIS offers several advantages for ISAC. Firstly, as shown in Fig. \ref{ISAC_model}a, STAR-RIS can enable full-space sensing and communication, which is a straightforward benefit. Secondly, STAR-RIS opens up new possibilities for carrying out sensing and communication in physically separated spaces using a shared hardware platform and waveform \cite{wang2022stars}, as illustrated in Fig. \ref{ISAC_model}b. Specifically, the STAR-RIS can divide the full space into two halves: the \emph{sensing half-space} and the \emph{communication half-space}. The ISAC node transmits a joint signal, which is then split by the STAR-RIS into two separate signals, for carrying out sensing and communication in the respective half-spaces. 

Nevertheless, STAR-RIS also poses new challenges. The transmission and reflection properties of the STAR-RIS can result in mixed echo signals from the two half-spaces in the link between the STAR-RIS and the ISAC node, making it difficult to identify the signals. Additionally, the multiple hops of the echo signal and energy splitting at the STAR-RIS can result in a low sensing SNR, leading to low sensing accuracy. To address these challenges, \cite{wang2022stars} proposed to install dedicated sensing elements at the STAR-RIS, which is referred to as the \emph{sensing-at-STAR-RIS} structure. Target sensing is carried out at these dedicated sensing elements instead of the ISAC node. Such a structure avoids mixing the echo signals in a single link and establishes a one-hop echo link, resulting in improved sensing accuracy. 

In Fig.~\ref{ISAC}, we evaluated the effectiveness of the \emph{sensing-at-STAR-RIS} structure, where the entire space is divided into the \emph{sensing half-space} and the \emph{communication half-space} by the STAR-RIS. The root Cramér–Rao bound (CRB) achieved for target sensing under varying communication signal-to-interference-plus-noise ratio (SINR) requirements is depicted in Fig.~\ref{ISAC}a. The CRB represents a fundamental lower bound for the mean square error achievable by any unbiased estimator. It is evident from the results that both independent and coupled phase-shift STAR-RISs exhibit superior performance compared to the conventional RIS in the ISAC system. The significance of the STAR-RIS in this context is further supported by practical estimation outcomes, as presented in Fig.~\ref{ISAC}b. Specifically, the estimation results obtained by the STAR-RIS exhibit a much smaller bright region than those obtained by the conventional RIS, thereby indicating a higher level of estimation accuracy.

\begin{figure}[t!]
    \centering
    \subfigure{
        \label{THz_SE}
        \includegraphics[width=0.8\linewidth]{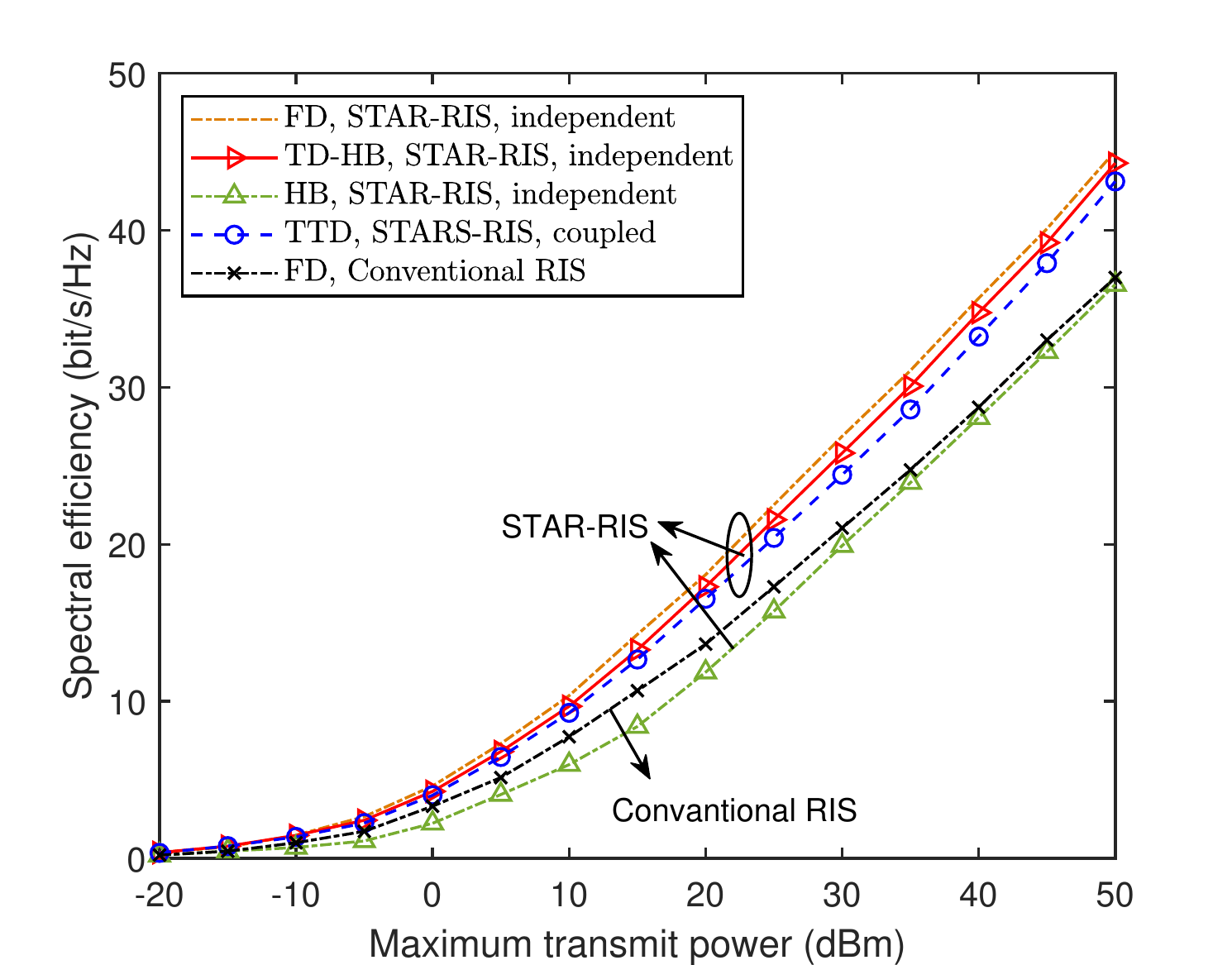}
    }
    \subfigure{
        \label{THz_EE}
        \includegraphics[width=0.8\linewidth]{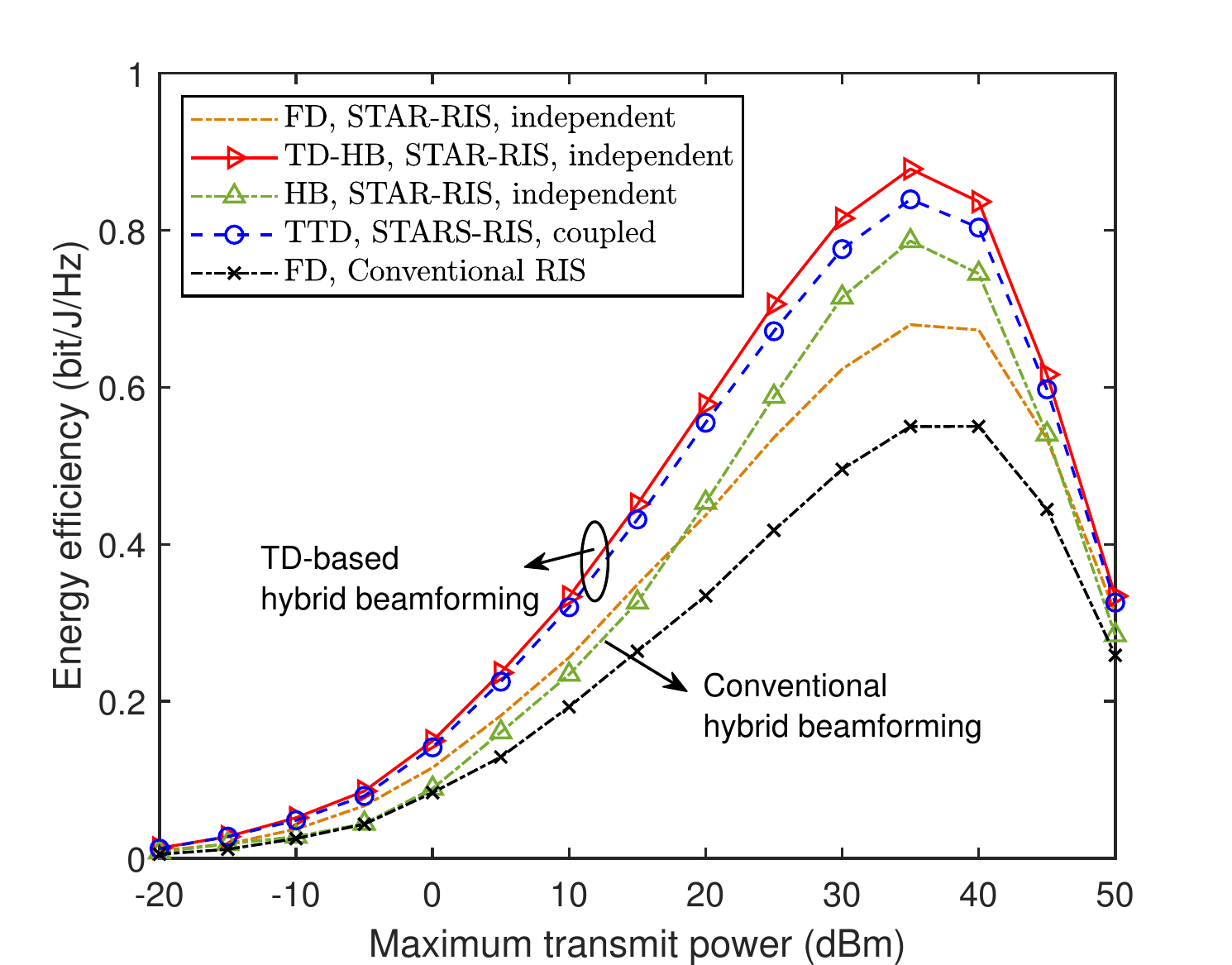}
    }
    \caption{Performance of the wideband STAR-RIS-THz system achieved by different beamforming structures in terms of a) spectral efficiency and b) energy efficiency. In particular, “FD”, “TD-HB”, and “HB” represents fully-digital beamforming, TD-based hybrid beamforming, and conventional hybrid beamforming, respectively. The detailed simulation setup can be found in \cite{wang2022simultaneously}.}
    \label{THz}
\end{figure}

\subsection{STAR-RIS-THz}
The vast bandwidth resources available in the THz band, which can reach tens of gigahertz (GHz), make THz communication a promising technique to support peak data rates of terabits per second (Tbps) in 6G wireless networks. However, THz signals experience significant propagation loss and high susceptibility to blockage due to their very high frequency. RIS can mitigate these issues by establishing additional line-of-sight (LOS) paths, but their reflecting-only property limits design flexibility. In contrast, the $\text{360}^\circ$ coverage capability of STAR-RIS technology makes it easier to address signal blockage. Furthermore, STAR-RIS can also bridge physically separated spaces, such as indoor and outdoor areas, which is not possible with reflecting-only RIS. 

To effectively exploit the bandwidth resources available in the THz band, it is imperative to consider the wideband STAR-RIS-THz system. However, the wireless channel can be significantly frequency-dependent, owing to the large array of the STAR-RIS and the very high carrier frequency used. This can induce a beam split effect, resulting in an array gain loss in the wideband STAR-RIS-THz system. The beam split effect can be directly caused by either phase-shifter (PS) based analog beamforming at the BS \cite{9398864} or passive beamforming at the STAR-RIS \cite{yan2022beamforming} due to their frequency-independent property. Recent research contributions have demonstrated that the beam split effect at the RIS depends on the size and shape of the RIS \cite{yan2022beamforming}. Therefore, it is possible that the two sides of the STAR-RIS  experience varying degrees of beam split effect, especially when employing the mode-switching protocol. To mitigate the beam split effect in the wideband STAR-RIS-THz system, \cite{wang2022simultaneously} proposed to introduce a time-delay (TD) network into the conventional hybrid beamforming structure. Since the TD is \emph{frequency-dependent}, such a TD-based hybrid beamforming can effectively reduce the beam split effect.

The performances of the wideband STAR-RIS-THz system in terms of maximum spectral efficiency and maximum energy efficiency via different beamforming structures are presented in Fig.~\ref{THz}. The results indicate a substantial performance improvement achieved by STAR-RIS over the conventional RIS in both cases. Furthermore, as depicted in Fig.~\ref{THz}a, the conventional hybrid beamforming structure leads to a significant spectral efficiency loss due to the beam split effect. Nevertheless, the introduction of the TD network enables the achievement of comparable spectral efficiency to that achieved through fully-digital beamforming. Furthermore, due to the relatively low power consumption of the TD network, it also helps to enhance the energy efficiency of the wideband STAR-RIS THz system, as depicted in Fig.~\ref{THz}b.

\begin{figure}[t!]
    \centering
    \subfigure{
        \label{PLS_external}
        \includegraphics[width=0.8\linewidth]{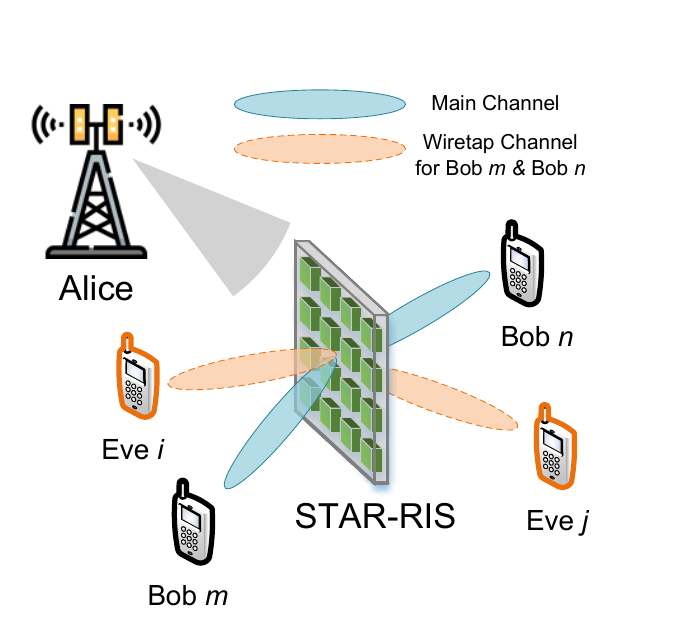}
    }
    \subfigure{
        \label{PLS_internal}
        \includegraphics[width=0.8\linewidth]{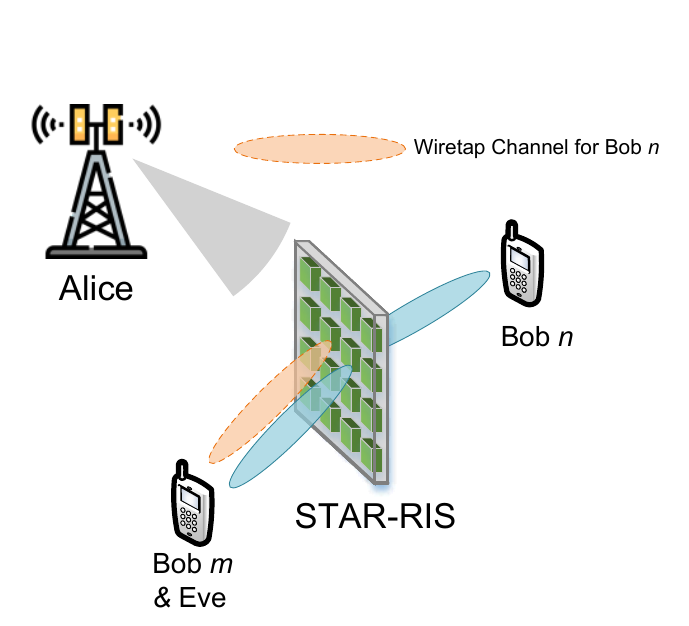}
    }
    \caption{Illustration of PLS of STAR-RIS with a) external eavesdropper and b) internal eavesdropper.}
    \label{PLS}
\end{figure}

\subsection{STAR-RIS-PLS}

Compared to the reflecting only RIS, the appearance of STAR-RIS provides extended coverage from $\text{180}^\circ$ to $\text{360}^\circ$. Nevertheless, such changes inevitably result in full-space wiretapping. As a result, the PLS issue needs to be re-investigated. As the basic model of STAR-RIS consists of both transmitted and reflected users, we can broadly classify the PLS problems into a couple of categories, namely, external eavesdropper case and internal eavesdropper case. 1) For the external eavesdropper case, as shown in Fig.~\ref{PLS}a, a few eavesdroppers try to eavesdrop on both transmit and reflect users. \cite{10040906} proposed a secure beamforming design for the coupled-phase shift STAR-RIS networks by maximizing the security capacity. We demonstrated the security performance advantages of our proposed design over conventional RIS, by only utilizing a few discrete phase shifters. 
2) For the Internal Eavesdropper case, due to the multi-user nature of STAR-RIS, internal eavesdropping may exist. More particularly, for a pair of transmitted and reflected users, one transmitted/reflected user may wiretap the other reflect/transmit user’s confidential information, which is regarded as STAR-RIS networks in presence of an untrusted user, as shown in Fig.~\ref{PLS}b. Such cases pose new challenges for the PLS in STAR-RIS networks and therefore deserve further research efforts. Furthermore, how to invoke artificial noise to further enhance the PLS for both external and internal eavesdropping cases is another promising direction to carry out.

\section{Conclusions and Future Directions}

This article provided a comprehensive overview of STAR-RISs for future 6G wireless networks. More particularly, we began by identifying three different models as fundamentals of STAR-RISs. We then summarized three representative categorizing approaches of STAR-RISs. Moreover, we investigated both the independent and coupled phase shift beamforming design of STAR-RISs. We demonstrated that our proposed generalized optimization framework can be exploited as an ``add-on'' approach to achieve near-optimal performance. This is followed by the discussion of several promising applications of STAR-RISs in future 6G networks. There are still numerous open research future directions in this area, some of which are listed as follows.
\begin{itemize}
\item \textbf{STAR-RIS Assisted Near-Field Communications}: In order to meet the demanding criteria for spectral efficiency, energy efficiency, latency, coverage, and other performance metrics in 6G, STAR-RIS follows primary developmental trends. The first involves the utilization of extremely large surfaces, while the second focus on the use of high-frequency bands, such as millimetre-wave (mmWave) and THz bands. These two trends will inevitably cause the near-field effect. More particularly, the \emph{Rayleigh distance}, which is known as the boundary between the near-field and the far-field, is proportional to the aperture of the antenna array and the operating frequency. An illustrative example of the challenges posed by the near-field effect can be observed in the case of a surface with a dimension of 0.5 meters operating at a frequency of 60 GHz. In this scenario, the near-field region can extend up to 100 meters, signifying that the impact of the near-field effect cannot be disregarded. In comparison to the conventional reflecting-only RIS, which features a fixed near-field region due to the fixed size and shape of the surface, the STAR-RIS offers a novel approach to dynamically adjust the near-field region through the allocation of distinct elements for transmission and reflection. Hence, it is imperative to consider the impact of the near-field effect in the development of advanced STAR-RIS-aided systems.

\item \textbf{Caching at STAR-RISs}: Recently, the design of multi-functional STAR-RIS becomes a promising research direction. In addition to the aforementioned active STAR-RISs and ISAC-STAR-RIS, caching at STAR-RIS can be also a promising attempt. More particularly, in contrast to conventional edge caching, caching at STAR-RIS is capable of 1) Improving network efficiency: By caching frequently accessed information at STAR-RIS, information can be quickly retrieved by nearby users without having to travel all the way to  the central server or edge nodes; 2) Achieving enhanced diversity: STAR-RIS adaptively transmits signals based on network caching information, by doing so, a couple of information from both BS and STAR-RIS may be obtained at receivers, for achieving the enhanced diversity. Nevertheless, the design of caching at STAR-RIS  requires additional active elements/relays to deploy on STAR-RIS, which brings the increased hardware cost. Moreover, as the caching processes consist of both information pushing  and  information delivery phases,  how to propose  efficient protocols for coordinating a proper design for different phases among a BS, active elements/relays and STAR-RIS is a non-trivial problem, which requires further research efforts.

\item \textbf{Stochastic Geometry  for STAR-RIS Assisted Networks}: Stochastic geometry is recognized as a powerful mathematical tool to capture the spatial randomness of wireless networks. It can be used to evaluate the average performance of one particular area. For example, we can invoke stochastic geometry to evaluate how many STAR-RISs  we need in a cafe, a campus, or even in a city. For the conventional reflecting-only or transmitting-only RIS, we may need to use Poission Hardcore processes to capture the characteristics of spatial deployment. This is because only half-space points can be active. This assumption is not technically friendly for stochastic geometry. Such an issue can be fully avoided by STAR-RIS due to its nature of providing $\text{360}^\circ$ full-space converge. As a result, the commonly used Homogeneous Poisson Point Process  can be invoked to provide spatial analysis, which can significantly simplify the mathematical analysis complexity. This characteristic brings new research opportunities for STAR-RIS networks. Of course,  advanced and suitable Point Processes are required to be further investigated, for providing accurate and efficient analysis.
\end{itemize}

\bibliographystyle{IEEEtran}
\bibliography{bibsample}

\end{document}